\documentclass[twocolumn,showpacs,preprintnumbers,amsmath,amssymb]{revtex4}
\usepackage{dcolumn}% Align table columns on decimal point
\usepackage[dvipsnames]{xcolor} %colors
\usepackage[utf8x]{inputenc}
\usepackage{graphics}
\usepackage{epsfig}
\usepackage{subfigure}
\usepackage{rotating}
\usepackage{mathrsfs}
\usepackage{amsfonts}
\usepackage{amsmath}
\usepackage{amssymb}
\usepackage{placeins}
\usepackage{comment}
\usepackage{bm}
\usepackage{pgfplots}
\usepackage{soul}
\usepackage{fixmath}
\usepackage{physics}
\usepackage[colorlinks]{hyperref}
\usepackage{ulem}

\begin{document}
\title{Multipartite quantum correlated bright frequency combs}

%Unveiling symmetries of multimode dynamics in bright SiN microcombs
\author{A. Bensemhoun$^{1}$, S. Cassina$^{2}$, C. Gonzalez-Arciniegas$^{3}$, M. F. Melalkia$^{1}$\footnote{Present address: Laboratoire Syst\`emes Lasers, Ecole Militaire Polytechnique, BP 17 Bordj El Bahri, 16046 Algiers, Algeria}, G. Patera$^{5}$,  J. Faugier-Tovar$^{6}$, Q. Wilmart$^{6}$, S. Olivier$^{6}$, A. Zavatta$^7$, A. Martin$^{1}$, J. Etesse$^{1}$, L. Labonté$^{1}$, O. Pfister$^{3,4}$, V. D'Auria$^{1}$\footnote{Contact author: virginia.dauria@univ-cotedazur.fr}, and S. Tanzilli$^{1}$}

 \affiliation{$^{1}$Universit\'e C\^ote d'Azur, CNRS, Institut de Physique de Nice, 17 rue Julien Laupr\^etre, 06200 Nice, France.}
\affiliation{$^2$ University of Insubria, Department of Science and High Technology, via Valleggio 11, 22100 Como, Italy}
\affiliation{$^{3}$\textcolor{black}{University of Virginia Physics Department, 382 McCormick Rd, Charlottesville, VA 22903, USA.}}
\affiliation{$^4$ Charles L. Brown Department of Electrical and Computer Engineering, University of Virginia, 351 McCormick Road, Charlottesville, VA 22903, USA}
\affiliation{$^{5}$University of Lille, CNRS, UMR 8523 -- PhLAM -- Physique des Lasers Atomes et Mol\'{e}cules, F-59000 Lille, France.}
\affiliation{$^6$ CEA-LETI, Universit\'e Grenoble Alpes, F-38000 Grenoble, France}
\affiliation{$^7$Istituto Nazionale di Ottica (CNR-INO), CNR, Largo Enrico Fermi 6, 50125 Firenze, Italy}

\begin{abstract}
        This experimental work demonstrates multipartite quantum correlation in bright frequency combs out of a microresonator integrated on silicon nitride operating above its oscillation threshold. Multipartite features, going beyond so far reported two-mode correlation, naturally arise due to a cascade of non-linear optical processes, making a single-color laser pump sufficient to initiate their generation. Our results show the transition from two-mode to multipartite correlation, witnessed by noise reductions as low as $-2.5$\,dB and $-2$\,dB, respectively, compared to corresponding classical levels. A constant of the movement of the non-linear interaction Hamiltonian is identified and used to asses the multipartite behavior. Reported demonstrations pave the way to next generation on-chip multipartite sources for quantum technologies applications.
\end{abstract}
\maketitle

\section{Introduction}

Optical multipartite states that exploit continuous variable (CV) encoding are key resources for quantum information science~\cite{Fabre2020}, playing a crucial role in measurement-based optical quantum computing~\cite{Pfister2019, Asavanant2022}, multipartner secure quantum networks~\cite{Epping2017} and quantum metrology~\cite{Guo_2019,Zhang2024}. 
Future ambitious applications will require advanced quantum systems combining the ability to generate and/or manipulate entanglement over a high number of modes with devices' robustness and small footprints. Such scalability can be obtained by spatial~\cite{Zhang2020}, temporal~\cite{yokoyama2013,yoshikawa2016,Asavanant2019,Larsen2019}, or spectral~\cite{Pysher2011,roslund2014wavelength,chen2014experimental,barbosa2018hexapartite} multiplexing. Spectral multiplexing is particularly attractive as large numbers of entangled modes can be directly generated in a single non-linear optical source, while eschewing the overhead from sequential processing~\cite{Fabre2020}. In addition, thanks to their different colors, quantum correlated modes can be easily separated and addressed using standard optical components~\cite{Fabre2020}.

Another key property for multiple applications is the ability to produce bright quantum entangled beams. For example, in quantum sensing, the ultimate quantum noise floor of optical interferometry decreases with the average photon number of the probing laser~\cite{Caves1980}. When the maximum usable light power is reached for whatever reason (record-high in LIGO, living-tissue vulnerability in biomedical imaging, embarked-power limitation on a drone), one can resort to single- or multi-mode squeezed light to further lower the detection noise. In most cases, however, this requires to displace a low-power squeezed state by reflecting it off a weakly transmitting beam splitter, through which a powerful laser beam is transmitted~\cite{PARIS1996}. This arrangement is extremely wasteful of laser power and requires dedicated locking systems, making it evidently advantageous to generate bright squeezed light directly, with no displacement operations. More generally, working with bright quantum beams gives the possibility of implementing self-homodyne detections to measure the continuous variable properties of entangled modes and, in turn, to access the complete reconstruction of their states via quantum tomography~\cite{barbosa2013beyond, alfredo2024quantum}. Compared to standard homodyne, this technique does not need an external local oscillator beam at the signal frequency and is particularly convenient when multiple beams at different colors must be detected simultaneously.  

There have been extensive explorations of bright bi- and multi-partite quantum correlations and entanglement in free-space systems exploiting, in particular, second order non-linear processes ($\chi^{(2)}$)~\cite{Heidmann1987,Feng2004,Villar2005,Coelho2009,barbosa2018hexapartite}. In view of future practical applications, this experimental work aims at investigating the generation of bright multipartite states in miniaturized optical sources integrated on a silicon-based platform. 

Integrated photonics on silicon (Si) and silicon nitride (SiN) is a go-to solution to reduce the physical size of optical systems while benefiting from cascaded optical components on CMOS compatible platforms~\cite{Chembo2016, Madsen2022, Labonte2024,zhang2021squeezed}. Frequency combs generated by four wave mixing (FWM) in silicon-based microresonators (microcombs) advantageously bridge multipartite entanglement and integrated resources~\cite{Chembo2010, Gouzien2023, Guidry2023, sloan2024}. \textcolor{black}{In below threshold devices pumped with a monochromatic input laser, the quantum state of low power microcombs has been proven to be made of many bipartite states independent of each other~\cite{vaidya2020broadband, yang2021squeezed, jahanbozorgi2023generation}. The use of multi-color input pumps has recently proved to engineer such bipartite states into multi-partite CV correlations~\cite{wang2024cluster, Jia2025}}. Bright frequency combs are efficiently obtained from above threshold microrings and disks~\cite{karpov2019dynamics}. Nevertheless, the complicated dynamics of the system, including the presence of bistability~\cite{Godey2014} and of morphing behaviour~\cite{Gouzien2023}, makes it non easy to transpose concepts proven in bulk systems into integrated ones ruled by FWM. \textcolor{black}{As a matter of fact, reported works only show CV quantum correlation between \textit{two} frequency modes out of microresonators pumped with a single input laser~\cite{Dutt2015, Dutt2024, alfredo2024quantum}}. A signature of multimode features has been measured from second order photon correlation in solitons out of systems far above threshold~\cite{guidry2022quantum}.
\newline
This paper goes beyond two-color analysis to demonstrate experimentally that a multipartite behavior is naturally present in bright microcombs from above-threshold SiN-microrings. \textcolor{black}{Remarkably, the generation of the multipartite state is initiated by a single monochromatic pump at the device input thanks to a cascade of FWM processes. We predicted this behavior in a previous theoretical work~\cite{bensemhoun2024multipartite}}. CV multipartite dynamics is explored here by measuring quantum intensity correlation between up to four bright optical modes generated by an SiN microring. The analysis of intensity correlation makes it possible to observe that, when driving the system progressively farther above threshold, a transition takes place from a bipartite regime to a multipartite one at four modes and beyond. \textcolor{black}{In our work, we go on to demonstrate a fundamental structure to these quantum correlations, epitomized by a multipartite phase invariant that we discover for any FWM interaction. This new quantum physics was never observed in microresonators operating below threshold ~\cite{wang2024cluster, Jia2025} because the cascade of FWM processes requires combs with bright frequency components, nor in above threshold works~\cite{Dutt2015, Dutt2024, alfredo2024quantum} as its experimental evidence demands the investigation of more than two modes. Our results pave the way to a new operation regime for quantum microcombs as they push bright correlation in microresonators beyond one-to-one schemes reported so far~\cite{Dutt2015, Dutt2024, alfredo2024quantum}. By doing so, they explore their potential for high-dimensional state connectivity and its applications in quantum photonics.}

\bigskip

\begin{figure}
\centering
\includegraphics[width=0.85\linewidth]{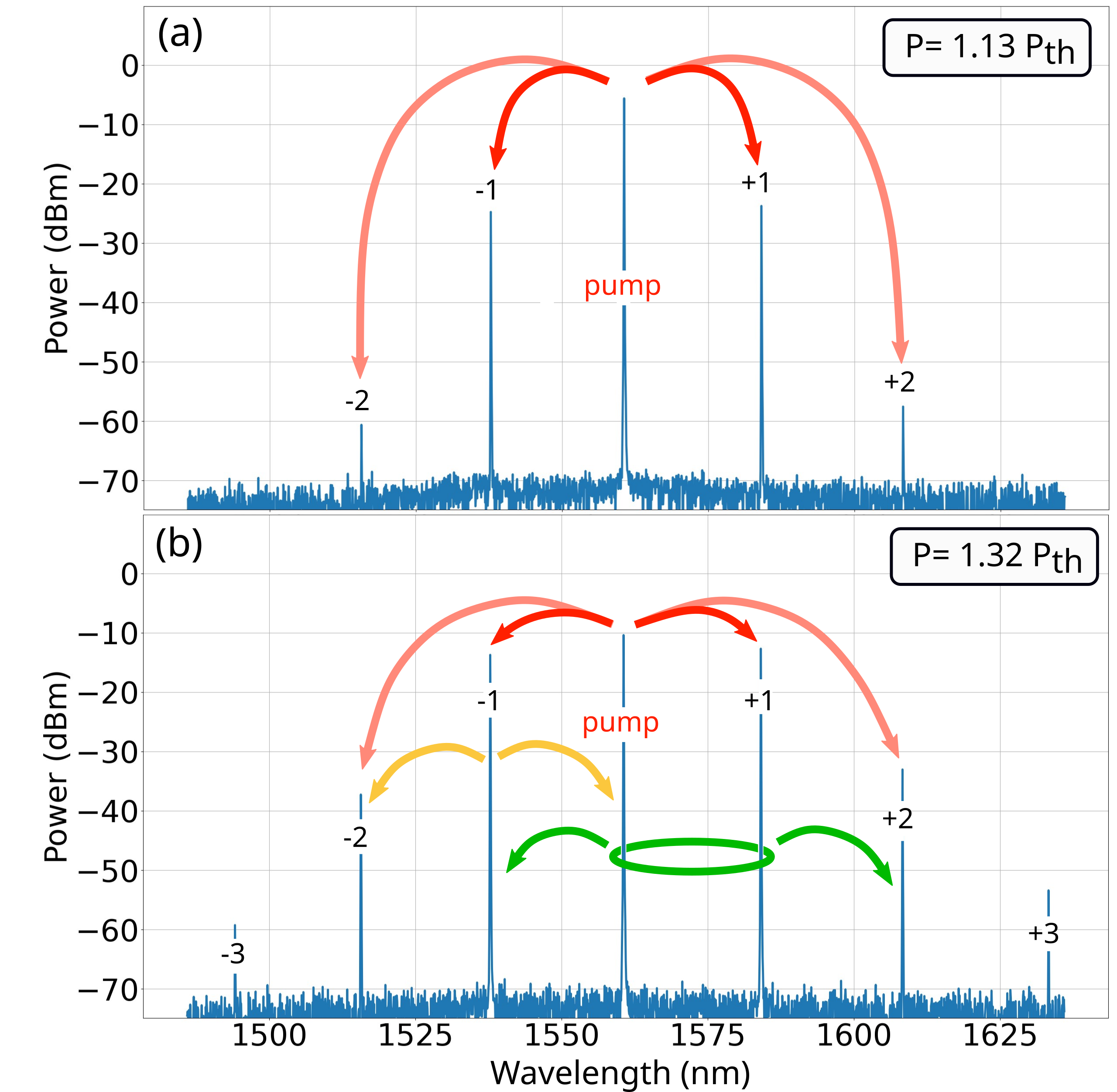}
\caption{Measured optical spectra of the comb generated by the SiN microring operating above the oscillation threshold for input pump powers of (a) $P=1.13~P_{th}$ and (b) $P=1.32~P_{th}$, where $P_{th}$ is the pump corresponding to the threshold. The figure shows some of the multiple non-linear processes that can simultaneously take place, including degenerate FWM of the input pump (red) or of a secondary mode (yellow), as well as non-degenerate FWM of two of the comb modes (green). Self- and cross-phase modulations are not shown to preserve the drawing legibility.}
\label{DescrizioneModi}
\end{figure}

\section{Multipartite correlation in bright microcombs}
\subsection{Concepts and experimental setup}
In above threshold FWM, the generation of bright frequency combs is associated with rich dynamics~\cite{karpov2019dynamics, Chembo2016, Guidry2023, Godey2014}. Figure~\ref{DescrizioneModi} shows some exemples of primary microcombs out of the SiN microresonator, measured by an optical spectrum analyzer (OSA). The system is pumped with a continous wave (CW) input laser and operates at moderate distance from the threshold, well below the soliton regime~\cite{karpov2019dynamics}; as shown in the figure, bright modes, including the input pump, are all in the telecom c-band. In these conditions, different FWM processes can take place~\cite{bensemhoun2024multipartite}. The single CW laser generates via degenerate FWM a primary comb of paired modes, symmetrically distributed with respect to the pump line (see modes $\pm 1$, $\pm2$ in Figure~\ref{DescrizioneModi}-a). Such bright components provide, in turn, additional pumps for other degenerate or non-degenerate FWM processes (see Figure~\ref{DescrizioneModi}-b). The coexistence of such simultaneous FWMs makes multipartite physics arise in a natural way when increasing the modes' powers. The higher is the input pump power, $P$, the more number of modes, as well as their relative power, increase making their contribution to secondary interactions non-negligible anymore. \textcolor{black}{A formal description of this behavior is provided in the analytical model described in Ref.~\cite{bensemhoun2024multipartite}.}

\begin{figure}
\centering
\includegraphics[width=0.85\linewidth]{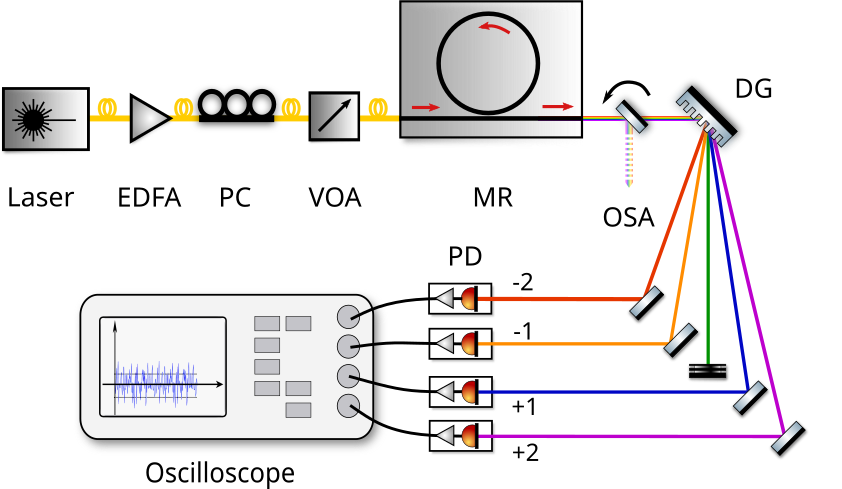}
\caption{Experimental setup for the measurement of intensity correlations at the output of an SiN microresonator (MR) above threshold. Modes $\pm$1 and $\pm$2 are separated by a grating (DG) thanks to their different colors and their intensities are individually detected (PD). \textcolor{black}{Note that, in the actual experiment the diffraction grating works in transmission, and two mirrors for separated beams are required downstream of it.} The corresponding photocurrents are evaluated in the Fourier transform domain and digitally combined as required to evaluate the correlation among different modes. Noise power spectrum of combined signals are used to retrieve non-classical features. Noise are evaluated with respect to shot noise signals obtained by sending to the detectors laser beams whose powers match those of the modes under scrutiny.} 
\label{Setup}
\end{figure}

The experimental setup used to generate and to investigate the multipartite state is sketched in Figure~\ref{Setup}. 
Before the microresonator, the setup is entirely made of plug-n-play telecom components based on optical fibers. The master source is a fiber-coupled CW laser (RIO-Orion) delivering 12\,mW at 1560.053\,nm \textcolor{black}{with a linewidth of $\leq 5$\,kHz: its central wavelength can be tuned of $\pm$20\,pm with minimal steps of 0.02\,pm.} The laser output is amplified by a Erbium Doped Fiber Amplifier (EDFA), whose working point is optimized to reduce noise contributions. Downstream of the EDFA, \textcolor{black}{residual amplified spontaneous emission is suppressed by a cascade of 5 wavelength demultiplexing stages (not represented), each providing a noise attenuation of -25\,dB and centered around 1560\,nm with a width of 100\,GHz, well above the laser tunability range}. After an in-line variable optical attenuator (VOA), the laser polarization is adjusted (PC) to match the transverse electric (TE) mode of the microresonator, whose propagation losses were measured to be the lowest. The laser light is edge-coupled via a microlensed fiber into the microresonator and used as the input pump of the FWM process. Fiber-to-chip coupling losses are of about -3\,dB. 
The microresonator is fabricated in-house at the CEA-LETI on Si$_3$N$_4$~\cite{ElDirani19, SpieLETI}. It is a single-pass microring (MR, see Figure~\ref{Setup}) with a radius of 112\,$\mu$m (corresponding to a free spectral range of 200\,GHz) coupled to a straight waveguide set at a gap of 300\,nm. The width and height of all waveguides are, respectively, 1.9\,$\mu$m and 800\,nm. The MR is \textcolor{black}{isolated against mechanical and thermal instabilities and} mounted on a thermo-electric cooler (TEC) that stabilizes its temperature at 68.5$^{\circ}$C. The microresonator has an intrinsic $Q\sim$10$^7$\textcolor{black}{with a cavity linewidth of the order of $\sim$630\,MHz} and operates in the overcoupling regime, with a measured pump oscillation threshold $P_{\text{th}}=$53$\pm$3\,mW. \textcolor{black}{Note that as in most works on non-linear optical resonators pumped in CW regime~\cite{vaidya2020broadband, yang2021squeezed, jahanbozorgi2023generation, Dutt2015, Dutt2024, alfredo2024quantum}, the laser linewidth is much smaller than the cavity resonance: such a narrow pump laser guarantees, in absence of secondary FWM effects, the generation of one-to-one correlations among symmetrical modes ($\pm$1, $\pm$2,...)~\cite{Chembo2016, Gouzien2023}. For the scopes of our analysis, this allows us to be able to start from a bipartite regime and to follow its transformation when secondary FWM processes become non negligible}. \textcolor{black}{For each measurement point, the laser wavelength is adjusted to guarantee its perfect resonance with the cavity resonance labeled as k=0. No active locking to the cavity resonances is used on the pump laser.} 

Its output is collected in free space with a microscope objective. \textcolor{black}{Since the chip facets are not provided with antireflection coating, the beams transmission, $\eta_c$ is reduced due to the Fresnell losses, that give $\eta_F=(\frac{n_{SiN}-n_{air}}{n_{SiN}+n_{air}})^2\approx 0.90$, with $n_{SiN}$ and $n_{air}$, the SiN and air refractive indexes.} A low-loss diffraction grating splits the comb's optical components and sends them to separate photodiodes (PD). \textcolor{black}{The grating works in transmission and has a groove density of $\approx$1000\,grooves/mm. We measured for our grating an isolation between next modes' wavelengths is of 45\,dB and a transmission $\eta_g\gtrsim$0.94 for all modes of interest. A flip mirror before the grating is used to send the light to the OSA to check the MR's operation regime before any measurement. Downstream of the grating, each beam is sent to two dielectric mirrors and a lens that direct it toward one of the photodiode (not depicted in Fig.~\ref{Setup}): the overall transmission of these optical components is $\eta_{opt}\approx$0.97.} Four modes ($\pm$1 and $\pm$2) are simultaneously detected. \textcolor{black}{Each photodiode mounts a head with an active surface of diameter 300\,$\mu m$ and a responsitivity at the telecom wavelength of $1.1$\,A/W, corresponding to a measured detection efficiency $\eta_{pd}\simeq$\,0.88}. Overall optical losses per mode lead to a detection efficiency \textcolor{black}{$\eta=\eta_F\eta_g\eta_{opt}\eta_{pd}\approx$\,72\%}. To evaluate the intensity correlation, the PD photocurrents are acquired by a fast oscilloscope to be evaluated in the Fourier domain and digitally combined. Noise power spectra of combined signals are used to retrieve non-classical features. All noise spectra are evaluated with respect to shot noise levels (SNL) that are obtained by sending to the detectors four laser beams whose mean intensities match those of the modes under scrutiny. \textcolor{black}{Details on the data acquisition, digital treatement and calibration with respect to the SNL are reported in Appendix.}   %Typical emissions are spaced by 14 FSR or by 15 FSR, depending on the excited primary microcomb.
\subsection{Quantifying multipartite intensity correlation}
\textcolor{black}{While different witness exist for multipartite correlations between field quadratures~\cite{wang2024cluster, Jia2025, Fabre2020, bensemhoun2024multipartite}, a witness for multipartite intensity correlations is not present in the literature. We can, however, construct it by looking at the Hamiltonian symmetries.} The overall system dynamics is described by the general FWM Hamiltonian, $\hat{H}_{FWM}$. It contains all processes by which two photons in frequency modes $p$ and $q$ annihilate to create paired photons in modes $s$ and $r$~\cite{bensemhoun2024multipartite, Gouzien2023}: 
\begin{equation}
\label{hamilFWM}
    \hat{H}_{FWM}=\displaystyle\sum_{rspq}^{} \: \delta_{r+s,p+q} \: \hat{a}^{\dagger}_{r} \: \hat{a}^{\dagger}_{s} \: \hat{a}_{p} \: \hat{a}_{q},
\end{equation}  
where $\hat{a}_i$ and $\hat{a}_i^{\dagger}$ are the bosonic operators associated to intracavity modes $i=0$ (the pump), $\pm1, \pm2,...$ and satisfy $[\hat{a}_i,\hat{a}_j^{\dagger}]=\delta_{i,j}$ and $[\hat{a}_i ,\hat{a}_j]=0$. The sum over all modes describes the fact that, above threshold, any pair of frequency modes can play the role of pumps for another FWM process, provided the energy conservation, expressed by the Kronecker delta, is respected (\textit{i.e.} $\omega_r+\omega_s=\omega_p+\omega_q$, where $\omega$ is the mode optical frequency). Cross- and self-phase correlation terms are included in $\hat{H}_{FWM}$~\cite{bensemhoun2024multipartite}. The non-linear interaction coupling constant is set to 1 for simplicity. \textcolor{black}{As detailed in Ref.~\cite{bensemhoun2024multipartite}, the Hamiltonian $\hat{H}_{FWM}$ allows constructing a set of linearized Langevin equations that link the quantum dynamics of each $\hat{a}_i$, to that of the other modes, via a coupling matrix whose elements depend on the classical amplitudes of the interacting fields, $\{A_i\}$: strong coupling are associated to high mean amplitudes. This formally translates the progressive appearance of multimode correlations when the fields at different frequency components become brighter.}
%Results of~\cite{bensemhoun2024multipartite} refer to quadrature entanglement. In the case of two modes, the noise figures for intensity differences are known to be proportional to those of quadrature differences~\cite{Fabre1989}, making the analysis of the analysis of ~\cite{bensemhoun2024multipartite} a solid theoretical basis for our work. 
\bigskip

Given $2M+1$ interacting modes, it is easy to prove that the Hamiltonian of Eq.~\eqref{hamilFWM} admits as a constant of motion the observable associated with the Hermitian operator:
\begin{align}
\label{Carlos}
\hat{\mathcal C}&=\displaystyle\sum_{k=-M}^{+M} \: k\, \hat{n}_{k} =\hat{n}_{1} -\hat{n}_{-1}+2\,(\hat{n}_{2} -\hat{n}_{-2} )+...,
\end{align}
with $\hat{n}_k= \hat{a}^{\dagger}_{k} \: \hat{a}_{k}$ and $k\in \mathbb{N}$. \textcolor{black}{In experiments, $\hat{\mathcal C}$ can directly be measured by simultaneously acquiring the intensities of all involved modes, $I_k \propto \hat{n}_k$, as functions of time.} The operator $\hat{\mathcal C}$ commutes with the time independent Hamiltonian $\hat{H}_{FWM}$, meaning that the associated physical quantity is conserved throughout the system evolution. In the case of a non-seeded microresonator, apart from the input pump ($k=0$), all modes $\pm k$ are initially in the vacuum state, leading to both $\langle \hat{\mathcal C} \rangle$ and Var($\hat{\mathcal C})=\langle \hat{\mathcal C}^2\rangle - \langle \hat{\mathcal C}\rangle^2$ equal to 0. These values are expected to remain constant even after the FWM interaction, when the modes are populated by bright beams. In particular, the variance of $\hat{\mathcal C}$ contains both quantum noise associated with individual modes and covariance terms describing intermodal quantum correlation. 
\textcolor{black}{It can be written as: \begin{equation}
\text{Var}(\hat{\mathcal C})= \text{Var}(\hat{\mathcal C}_{BP})+\text{Cov}(\hat{\mathcal C}_{MP}),
\label{CarlosVariance}
\end{equation}
where:
\begin{equation}
\begin{split}
\text{Var}(\hat{\mathcal C}_{BP})&=\displaystyle\sum_{k=0}^{+M} k^2\, \text{Var}(\hat{n}_{k}-\hat{n}_{-k})\\
\text{Cov}(\hat{\mathcal C}_{MP})&=\sum_{\substack{k,l=-M \\ k\neq l}}^{+M} k\,l\,(\langle \hat{n}_{k}\hat{n}_{l}\rangle-\langle\hat{n}_{k}\rangle\langle\hat{n}_{l}\rangle).
\label{varcov}
\end{split}
\end{equation}
In the previous expressions, $\text{Var}(\hat{\mathcal C}_{BP})$ covers all bipartite correlations between paired symmetric modes $\pm k$, while $\text{Cov}(\hat{\mathcal C}_{MP})$ contains all correlations between non symmetrics modes: such a covariance term is 0 for a state containing only bipartite correlations among symmetric modes, while it is in general non-negligible for multipartite systems.}
In the following, a $\text{Var(}\hat{\mathcal C}\text{)}$ below the shot noise level is used to witness quantum intensity correlation involving all interacting modes. This choice allows introducing a measure of multipartite correlation in intensity, differing from other common correlation witnesses that are based on field quadratures~\cite{Fabre2020}. Also note that $\hat{\mathcal C}$ represents one of the symmetries of the FWM Hamiltonian. Its associated evolution operator, $\mathcal{U}(\theta)=\prod_{k=-M}^M e^{i \theta k \hat{n}_k}$, is the operator that simultaneously phase shifts all modes $k$, each by an angle $ k\theta$. $H$ is invariant under $\mathcal U(\theta)$, $\forall \theta$.\newline

\section{Results}
\begin{figure}
\centering
\includegraphics[width=0.85\linewidth]{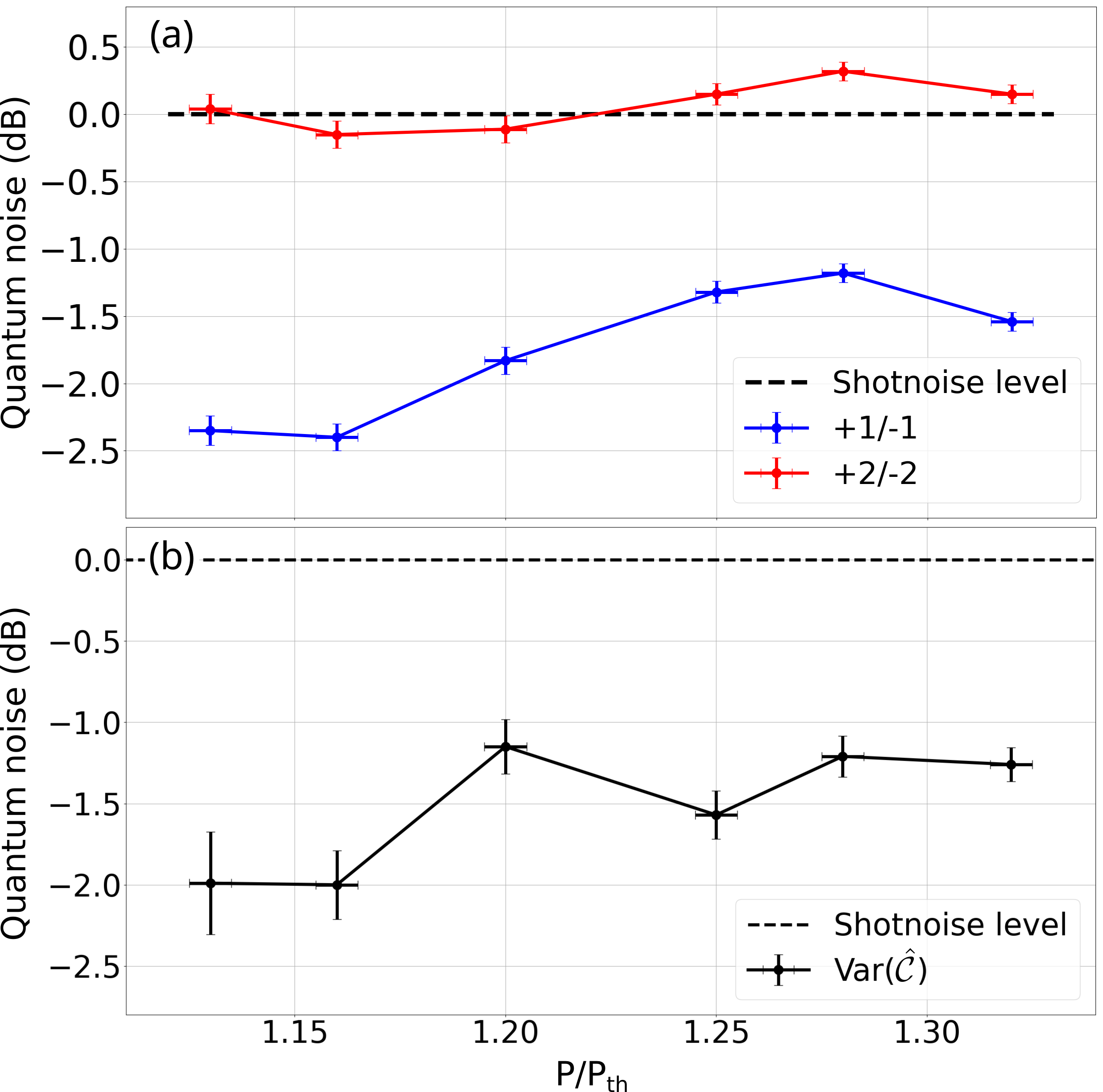}
\caption{(a) Quantum correlation between +1/-1 (blue) and +2/-2 (red) normalized to the shotnoise level as a function of the input pump power \textcolor{black}{divided by the oscillation threshold pump, $P_{\text{th}}$}. (b) Quantum noise normalized on the shotnoise level of the multipartite observable as a function of the input pump power. Experimental data are not corrected by quantum efficiency, only for residual electronic noise. All values refer to noise level evaluated at the same analysis frequency ($\approx$ 4MHz). For each of the curves the shot noise has been experimentally obtained by replacing each of the beams under scrutiny with a coherent beam of the same mean power. \textcolor{black}{In all the figures, $P$ is the input power coupled to the microresonator.}}
\label{CarlosMisurato}
\end{figure}

 Figure~\ref{CarlosMisurato} shows intensity correlation between modes $\pm$1 (a-blue), modes $\pm$2 (a-red) and the variance of $\hat{\mathcal C}$ on the ensemble of the four modes (b). Curves are plotted a functions of the input pump power, $P$, normalized to $P_{\text{th}}$. Bimodal intensity correlation are expressed in terms of the noise on their intensity difference that is proportional to Var($\hat{n}_{k}-\hat{n}_{-k})$. This approach has been already adopted to study modes $\pm$1 in bright microcombs~\cite{Dutt2015, Dutt2024}.

At low pump powers, the leading non-linear process is the degenerate FWM of the input laser beam that produces symmetric modes (see Figure~\ref{DescrizioneModi}-a). This leads to one-to-one correlation between modes $\pm$1, that are, to first approximation, independent of $\pm$2. This regime corresponds to the optimal intensity correlation of -2.5\,dB at low $P$/$P_{\text{th}}$ plotted in Figure~\ref{CarlosMisurato}-a. When increasing $P$, the intensity correlation of modes $\pm$1 is degraded, eventually reaching $\approx$-1\,dB. Such a change can be associated with the appearance of cascaded FWM processes that link the dynamics of $\pm$1 to that of $\pm$2, and then to that of $\pm$3, and so on to higher-order modes~\cite{bensemhoun2024multipartite}. Under these conditions, multipartite correlation arise and detecting only two modes corresponds to tracing out a part of the multipartite system; the measurement of Var($\hat{n}_{1}-\hat{n}_{-1}$) thus becomes insufficient to fully characterize intensity correlation available at the microring output. \textcolor{black}{This behavior is consistent theoretical results from two-mode correlations~\cite{bensemhoun2024multipartite}.} \textcolor{black}{Figure~\ref{DescrizioneModi}-a) also shows the correlation between modes $\pm$2, never observed so far in experiments~\cite{Dutt2015, alfredo2024quantum}. Although measured noise reductions are lower than for modes $\pm$1, also intensity correlations in modes $\pm$2 show a decrease of correlation with the increase of the pump power}. Lesser noise reductions in such modes are due to reduced signal-to-noise ratio when measuring weak intensity signals.  

The measurement of $\hat{\mathcal C}$ combines these results for the ensemble of the 4 modes. As expected, Var($\hat{\mathcal C}$) $<$  SNL for all considered input pump powers, and it goes down to -2\,dB at low $P$. Remarkably, this stays true despite the fact that, for this microcomb, Var($\hat{n}_{2}-\hat{n}_{-2}$) stays high and, eventually, passes above its SNL. \textcolor{black}{Interestingly,
experimental results on Var($\hat{\mathcal C}$) can be compared with Var($\hat{\mathcal C}_{BP}$), \textit{i.e.} with noise levels that are expected for $\hat{\mathcal C}$ when the microcomb quantum state contains \textit{only} bipartite correlations (see Eq.~\eqref{CarlosMisurato}). For each P/Pth, these can be computed by injecting in Eq.~\eqref{varcov} measured Var($\hat{n}_{1}-\hat{n}_{-1}$) and Var($\hat{n}_{2}-\hat{n}_{-2}$). Obtained Var($\hat{\mathcal C}_{BP})$ evolve from -1.6\,dB at $P/P_{th}\sim $1.1 to $\approx$+1\,dB when $P/P_{th}\geq$1.25 and are systematically higher than our measured Var($\hat{\mathcal C}$): the difference between Var($\hat{\mathcal C}_{BP}$) and Var($\hat{\mathcal C}$) confirm the presence of non-negligible covariance terms Cov($\hat{\mathcal C}_{MP}$) in Eq.~\eqref{CarlosMisurato} and, in turns, of cross correlations among non-symmetric modes. This quantum feature was out of the reach of former experimental works on bright microcombs that were restricted to the analysis of modes $\pm$1~\cite{Dutt2015, Dutt2024, alfredo2024quantum}.}

For the microcomb of Figure~\ref{CarlosMisurato}-b, an abrupt decrease of $\text{Var(}\hat{\mathcal C}\text{)}$ can be observed \textcolor{black}{when $P/P_{\text{th}}$ goes from 1.15 to 1.20 (from -2\,dB to $\approx$-1\,dB)}. Correspondingly, modes $\pm$3 start being non negligible in the microcombs. Although weaker than modes $\pm 1$ (typically -40\,dB), $\pm3$ terms act as seeds to secondary FWM processes, thus connecting their dynamics to that of other modes. In this regard, note that the measurement of $\hat{\mathcal C}$ with $M$=4 neglects contributions from symmetric modes $\pm3$ and higher; \textcolor{black}{disregarding them is equivalent to trace-out information on a part of an entangled state of higher dimension. The increase of Var($\hat{\mathcal C}$) observed when $P/P_{\text{th}} \geq1.20$, thus gives an indirect proof of a richer dynamics where more than 4 modes are quantum correlated.} 

\begin{figure}
\centering
\includegraphics[width=0.8\linewidth]{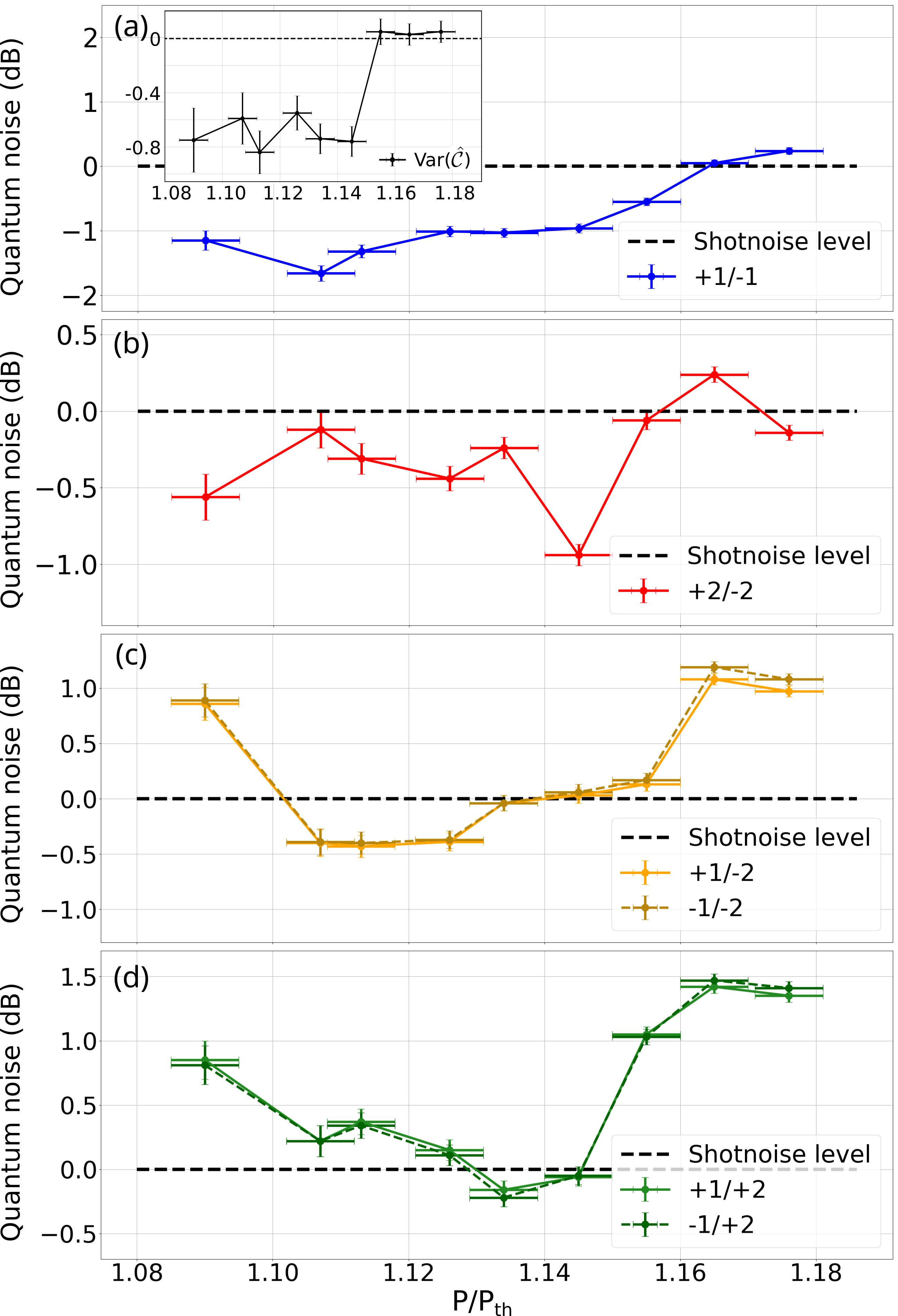}
\caption{Quantum correlation normalized to SNL as functions of the input pump power for modes +1/-1 (a), +2/-2 (b) and between non-symmetric modes (c) and (d). The inset reports the corresponding Var($\hat{\mathcal C}$).}
\label{cross_lowP}
\end{figure}

To dig into the four-mode regime, a second set of measurements is performed at lower pump values, in a regime where modes $\pm 3$ are expected to be negligible. Note that the microcomb under study in the low power regime is not the same as for measurements in Figure~\ref{CarlosMisurato}. This is due to optical bistability in resonant FWM processes, \textit{i.e.}, to the existence of two stable stationary solutions in above-threshold microresonators~\cite{Godey2014}; depending on the chosen range of input powers, the system spontaneously emits one microcomb or the other. \textcolor{black}{As explained, the modes' coupling and, thus, the actual levels of their correlation depend on the field classical amplitudes, $\{A_i\}$ (with $A_i\in {\mathbb{C}}$) that are expected to be different from a comb to the other~\cite{bensemhoun2024multipartite}. Nevertheless the overall dynamic should follow a behavior analogous to the one of Figure~\ref{CarlosMisurato}.} This can be seen in Figures~\ref{cross_lowP}-a) and b), that show intensity correlation as functions of low $P/P_{\text{th}}$ for symmetric modes; the behavior of Var($\hat{n}_{k}-\hat{n}_{-k})$ qualitatively confirms what has already been observed for the microcomb of Figure~\ref{CarlosMisurato}.
Lower measured correlation levels are due to the microcomb emission that is not bright enough to make residual electronics noise negligible. Figures~\ref{cross_lowP}-c) and -d) illustrate the dynamics of cross correlation between non-symmetric modes in terms of Var($\hat{n}_{i}-\hat{n}_{j}$), with $i \neq j=\pm1,\pm2$. 
At very low powers, secondary FWM processes are not yet well established and quantum correlation only take place between symmetric modes generated by degenerate FWM of the input pump (Figure~\ref{cross_lowP}-a) and -b). The reduction of Var($\hat{n}_{i}-\hat{n}_{j}$) observed when increasing $P$ bears witness to the birth of multipartite correlation: the links between modes initially independent lead to noises below the classical level at intermediate values of $P$ (\ref{cross_lowP}-c) and -d)). This regime can be considered as showing four-partite features. At $P/P_\text{th}\approx1.15$, the growth of modes $\pm3$ reduces cross- as well as symmetric-mode correlation; this represents a confirmation of a 6-partite dynamics for which neglecting information of modes above $\pm$2 corresponds to tracing out a part of a larger quantum correlated system. \textcolor{black}{The differences between curves of Figure c) and d) can be justified by optical losses and dispersion. Dispersion, in particular, affects the position of cavity resonance and can induce some asymmetries, especially experienced by modes $+2$ and $-2$, that are the most spaced in terms of optical frequencies. At the same time, in CW regime, only modes that are distributed in a perfectly symmetric way with respect to the input pump can satisfy the energy conservation for the initial degenerate FWM process: the mismatch of mode positions and cavity resonances affects the dynamics of the system and modify the strength of quantum correlations between symmetric and asymmetric modes~\cite{bensemhoun2024multipartite}}.

As a final remark, note that the behavior of symmetric and cross mode correlation is well summarized by the Var($\hat{\mathcal C}$)  evolution reported in the figure inset. Its steep change at higher power reproduces that of cross correlation and leads to values above the SNL at high powers. For this microcomb, in addition to the appearance of modes $\pm3$, it is pertinent to note that the higher is $P$, the more probable becomes the transition towards the comb corresponding to higher power, thus potentially affecting the system quantum correlation.
\section{Conclusion} 
The presented experimental work demonstrates multipartite features in bright frequency combs out of an SiN microring exploited above its oscillation threshold. Multipartite features appear spontaneously due to a cascade of secondary FWM processes that exploit the primary comb out of a degenerate FWM of a monochromatic pump laser. The dynamics of quantum correlation in four bright modes is investigated by measuring their one-to-one correlation as well as in terms of the noise of the operator $\hat{\mathcal C}$ that represents a symmetry of the FWM Hamiltonian and that combines all modes. Although experimental data refer to the first four modes appearing in the primary combs, all obtained results indicate that higher order connectivity can be expected when including in the analysis subsequent components that appear when working farther away from the threshold. Experimental evidence of multipartite correlation in bright microcombs and the study of their dynamics have never been performed before. Presented results, as well as the theoretical and experimental tools dedicated to their measurement, paves the way to the exploitation of bright microcombs for future quantum technology applications in CV multipartite quantum photonics, with a simple experimental configuration, where just one external CW pump laser is needed to feed the system. \textcolor{black}{The experimental tools and concepts demonstrated in this work apply in a safe way to primary combs of any size. The actual number of components in the microcomb depends on the device geometry and performance and scales with the input pump~\cite{karpov2019dynamics}, until the generation of the secondary comb is reached, eventually leading to a soliton regime. A theoretical~\cite{Guidry2023} and experimental paper~\cite{guidry2022quantum} have already discussed the presence of quantum correlation among the soliton frequency components. This result, combined with ours, shows that no strict limitation to the size of the correlated states are present in the shift from the primary comb to the soliton. Nevertheless, compared to other systems~\cite{Labonte2024}, the correlations of modes is predetermined by the classical emission comb: in this sense, our work thus opens to new theoretical and experimental investigations, in which the analysis of correlation between the bright modes is conducted with the perspective of engineering the quantum correlations as done for instance in below threshold systems~\cite{Jia2025, wang2024cluster}. } 
\section*{Acknowledgment}
This work has been conducted within the framework of the project OPTIMAL granted by the European Union by means of the Fond Européen de développement régional (FEDER). The authors also acknowledge financial support from the Agence Nationale de la Recherche (ANR) through the projects SPHIFA (ANR-20-CE47-0012) and OQulus (ANR-22-PETQ-0013) and from Academy RISE of Universit\'e C\^ote d'Azur. VDA thanks the Institut Universitaire de France for the support. OP acknowledges support from CNRS, Fondation Doeblin, and US NSF grants ECCS-2219760 and PHY-2112867.
\section*{Competing interests}
The authors declare no competing interests.
\section*{Authors' contributions}
A.B. mounted the entire experimental setup and took care of data acquisition and analysis, with the help of S.C., M.F.M., and A.Z.; C.G.-A. found the original theoretical tool for witnessing multipartite intensity correlation, with the help of O.P.; S.O. and Q.W. designed and fabricated the SiN integrated structures; V.D’A. and O.P. conducted the data analysis and interpretation, with the help of A.M., J.E., G.P., L.L., and S.T. All authors read, discussed, and contributed to the writing, reviewing, and editing of the manuscript. V.D'A., L.L., and S.T. coordinated and managed the project, ensuring its successful completion.

\section*{Appendix: Data acquisition, treatment and normalization}
\textcolor{black}{Downstream of the grating the beams are simultaneously detected by 4 photodiodes  in-house built at INPHYNI and LENS-INO. The detection circuits allow to split the DC and AC components of photocurrents.
The DC outputs of the photodiodes are directly proportional to the mean optical power of input beams. They are sent to an oscilloscope with a low bandwidth of 250\,MHz, to monitor the individual powers of the 4 modes: these values are required to retrieve corresponding shot noise levels that will be used in the noise normalization. }
\textcolor{black}{The AC amplification circuit is in a transimpedance configuration. We took care of balancing the performances of detectors measuring symmetric modes. Accordingly for the pair of detectors dedicated to modes $\pm$1 have an AC bandwidth at -3 dB of 69\,MHz while the detectors for modes $\pm$2 have a AC bandwidth of 10\,MHz. The choice of the analysis frequency, 4\,MHz, is based on the individual photodiodes detection spectra and allows us to obtain the best signal-to-noise ratios with respect to the detectors’ noise floors. }

\textcolor{black}{The detectors' photocurrents are sent to the four channels of a fast oscilloscope with a bandwidth of 4\,GHz and acquired at a sampling rate of\,5 GSample/s: due to the involved bandwidths, to avoid the effect of fast electronic noise components, we used a low-pass filter option on the oscilloscope to cut off frequency above 20\,MHz. For each measurement point, we made 50 subsequent acquisitions of 100\,$\mu$s. Acquired data are then treated with Python. Their Fast Fourier transforms are computed with a Python FFT library, and then combined according to the desired observables. For each analysis frequency, $\Omega$, digital data processing is used to compensate possible power unbalance and temporal delays between photocurrents associated to the detection of symmetrical modes $\pm$k. This is done by evaluating the difference $\Delta_k(\Omega)=(I_k(\Omega)-\alpha_k(\Omega)I-_k(\Omega))$ with $I_j(\Omega)$ the component at frequency $\Omega$ of the noise spectrum of the photocurrent associated to the detection of mode “j”. $\alpha_k(\Omega)$ is a complex parameter that used for balancing the signals’ power balancing and synchronization. The mean value of $\alpha_k$ is given by the DC components of the photocurrents, that are directly proportional to the mean power of input beams. The phase of $\alpha_k(\Omega)$ account for delay in temporal domain and  in principle depends on $\Omega$; its value is found by asking the algorithm to minimize $\Delta_k(\Omega)$. In the measurement of the observable $\hat{\mathcal C}$, in addition to $\alpha_1$ and $\alpha_2$, that are determined from $\Delta_1(\Omega)$ and $\Delta_2(\Omega)$, a parameter $\beta(\Omega)$ is used to comply with the overall delay between the two pairs of photodiodes. Based on the chosen analysis frequency, we implement the procedure for the determination of the three parameters at $\Omega$= 4\,MHz.}
\textcolor{black}{Once the digital parameters are set, for each measurement point, the program makes an average of data obtained over the 50 subsequent acquisitions and then compute the variance of obtained noise spectra. As a final step, correlation spectra are normalized to assess noise reduction (or increase) with respect to the shot noise associated to coherent beams. Shot noise traces come from the non-amplified laser, whose noise features have been carefully measured to exclude the presence of undesired classical noises within the detection bandwidths of our photodiodes. We calibrated the shot-noise by sending on the photodiodes the laser and measuring the spectrum of noise of the corresponding photocurrent. Do to so, the photocurrents undergo the same acquisition and digital data processing as for the signal beams and their noise feautures in Fourier domain are acquire. The shot noise data are acquired as functions of the laser mean powers, and the calibration is performed independently for each photodiode. For each of investigated observables ($\hat{\mathcal C}$ or $\hat{n}_k-\hat{n}_{-k}$ ), we computed the analytic expression of the quantum noise that would be obtained by replacing each of the modes $\pm$1, $\pm$2, with a coherent beam of the same optical power. To retrieve its actual experimental value, we used in the obtained formulas the shot noise levels that we got from the calibration: the obtained result is used to normalized the noises on Var($\hat{\mathcal C}$) or on Var($\hat{n}_k-\hat{n}_{-k}$). }

\end{document}